# Context Awareness Framework based on Contextual Graph

Tam Van Nguyen, *Student Member, IEEE*, Wontaek Lim, Huy Anh Nguyen, and Deokjai Choi
Department of Computer Engineering, Chonnam National University, Korea
Email: vantam@gmail.com, rltk139@daum.net, anhhuy@gmail.com, dchoi@chonnam.ac.kr

*Abstract* - **Nowadays computing becomes increasingly mobile and pervasive. One of the important steps in pervasive computing is context-awareness. Context-aware pervasive systems rely on information about the context and user preferences to adapt their behavior. However, context-aware applications do not always behave as user's desire, and can cause users to feel dissatisfied with unexpected actions. To solve these problems, context-aware systems must provide mechanisms to adapt automatically when the context changes significantly. The interesting characteristic of context is its own behaviors which depend on various aspects of the surrounding contexts. This paper uses contextual graphs to solve the problem "the mutual relationships among the contexts". We describe the most relevant work in this area, as well as ongoing research on developing context-aware system for ubiquitous computing based on contextual graph. The usage of contextual graph in context-awareness is expected to make it effective for developers to develop various applications with the need of context reasoning.**

## I. INTRODUCTION

Recent years have witnessed the development of wireless networks and mobile devices in ubiquitous computing. In that trend, the computing are moving towards pervasive with devices, software agents, and services all expected to seamlessly integrate and cooperate in support of human objectives – anticipating needs, negotiating for service, acting on our behalf, and delivering services in an everywhere, every time fashion [1]. Context awareness concentrates on supporting automatic services in ubiquitous environment.

Nowadays, many researchers have focused on context-aware architecture and context-aware applications. Those approaches can be listed as systems based on rule based, case-based or decision tree. But those systems have some limits such as no consideration on mobile users or no learning mechanisms to acquired individual users' preferences or users' interests. Context which is never in static mode changes spontaneously while the current context computing is too demanding of attention; too isolating from other people and activities. In the specific environments, like home, office, classroom, and so on, the interesting thing is that the context regularly changes its state from the old state to the new one. Another interesting thing stays in the behaviors of people, the main factors causing the changes of context. The people behaviors are different from others to others. On the other words, in the same situation, a person's way is different from the others' ways. Using contextual graphs is an approach targeting problem resolution in domains where context information is represented in a graph. Contextual graphs aim to solve the problem "the mutual relationships among the contexts" which caused by the context mobility. Also, we propose a context awareness framework based on contextual graph in order that developers can develop various applications with the need of context reasoning easily.

The remainder of this paper is organized as the follows. In the Section II, we begin to discuss the little survey on related works and their shortcomings which do not make effective reasoning of context. In Section III and Section IV we present the proposed contextual graph and the context-awareness framework, respectively. Section V describes an application of applying context reasoning by using the proposed framework. The conclusions and future works are given in Section VI.

## II. RELATED WORKS

A number of approaches are proposed the context-aware framework to support ubiquitous computing. The first approach is based on rule base in which knowledge is gathered as rules. These rules are pieces of knowledge of the form "if preconditions then conclusions." They are recorded in large rule bases difficult to update. The rules are structured chunks of knowledge, which are easily understood. However, the lack of structure of the rule-base impedes the comprehension (even for the experts of the domain) and the maintenance of the knowledge. Some works have been done on rule-bases structuring, namely on the splitting of the rule bases into several rule packets, each containing a subset of rules applied to solve a specific sub-problem [2]. Clancey [3] proposed to add screening clauses to the precondition part of the rules so that they are activated only in some kind of context, this amount to add in the preconditions of the rules some clauses constraining the triggering to a certain context. This is burdensome because the designers must anticipate all the possible contexts to define the preconditions of the rules. Moreover each rule describes a part or a whole context without any reference to the dynamics of the situation which is fundamental to understand the situation.

The second approach is using decision trees to acquire the decision of context-awareness. W.Y. Lum et, al. use decision trees to decide the optimal content version for presentation, base on the specific context, such as intended target device capabilities, network conditions, and user preferences [4]. A. Ranganathan et, al. propose a use first order model to express complex rules, which involves contexts. H. Chen et, al. use ontology models to express basic concepts of people, agents, places, and presentation events in an intelligent meeting room environment [5]. It's convenient to integrate the agents that employing knowledge. It's also convenient to reason to understand the local context. The decision tree approach itself [6] tries to represent the decision step by step. This is obtained by the presence of two types of nodes: the event nodes and the decision nodes. At an event node, paths are separated according to all the possible realization of an event on which the decision maker has no influence. On a decision node, the person makes a choice. This approach might be a way to structure rule bases. For each new element analyzed in the preconditions, a new event node is created. For each new value of an existing contextual element, a new branch is created, and so on. Rule after rule, a tree is constructed. The leaves give the rule conclusions. However, the main problem with this structure is *the combinatorial explosion*. The number of leaves is an exponential function of the depth of the tree. The addition of a contextual element may easily double the size of the tree.

Another interesting approach is the Case-Based Reasoning (CBR), which is a kind of analogy reasoning [7]. CBR is an approach targeting problem resolution in domains where little information is known about the key processes and their interdependencies. For context aware applications, at the beginning, we don't know the interdependency among appliances and services. And also, there is no theory to identify the context situation. L.D. Xu et al. discuss the CBR's advantages and the process of the CBR and provide an application that uses CBR to judge the AIDS [8]. W.C. Chen et, al discuss the features that can delegate the case [9]. They propose a framework to mine the features by using machine learning methods. W. J. Yin et al. use the joint of genetic learning approach and case-based learning, solving job-shop scheduling problems [10]. The similarity calculation is defined based on DNA matching. D. Grosser et al. use CBR to predict the object oriented software's stability [11]. K. Li et al. introduce time function as the adjustable factors in similarity measuring [12]. But awareness can be obtained by retrieving and adapting the solutions to previous scenarios. The main advantage of this reasoning is its great power of generalization and its maintenance. However, it fails to provide explanations on the obtained solution.

### III. CONTEXTUAL MANAGEMENT

In this section, we propose the contextual graph to overcome the shortcomings of contextual management described in the previous section.

#### A. Context representation

Context is "any information that can be used to characterize the situation of an entity. An entity is a person, place or object that is considered relevant to an application including the user and the applications themselves" [13]. As mentioned above, context always changes every time. In the real environment, such as smart home, sometimes means automated home, the typical used case is as follows: "After working day, Mr. Kim enters the living room; the room temperature is $30^oC$, the air condition will automatically turn on to decrease the temperature. The TV is turned on and the news report channel is tuned at the same time. At night, Mrs. Kim leaves the living room and enters the bedroom. At this moment, the smart home does the action "Take a rest". The air condition and TV in the living room are turned off. Meanwhile, the light in bedroom is turned on, and adjusted the brightness to low." The purpose of context-awareness is to do right things in right situations. The basic of the common reasoning context is the understanding of current state. But there are many situations TV, air conditioner and light might happen. It almost means impossible for the system designer to envision all possible contexts prior to the system deployment. The home system will sometimes perform in unexpected and undesirable ways inevitably and thus disappoint the home occupant. A common learning algorithm also can't solve this problem because a training set will not contain examples of appropriate decisions for all possible contextual situations.

Our current goal is to provide a uniform and well-organized context model from which to control and adapt the behavior of the application to satisfy user circumstances. One interesting thing in context analysis is the context transition. When a context changes its state to a new one, that context doesn't change completely, just a part in the whole context. For instance, context C1 changes into context C2, doesn't mean all of parameters of C1 change significantly, just a small part of context C1 such as temperature or light changes in order to create the new context C2 instead. On the other words, the core part or the static part, which is called shared context, regularly changes. The discovered shared context between two individuals can lead to a fruitful exchange of knowledge between them and triggers intra-organizational communication. This paper concentrates on that context movement or contextual changes.

Before storing into the dataset, the context aware system categorizes the contexts according to the relevance to users and context parameters. There are many types of parameters, such as activity, location, time, device properties, and user preferences. For example, in the real-time system, if time parameter contains day, month, year, hour, minute, second, there is not enough space to store the context because the unlimited number of contexts. The number of context will be uncontrollable. Instead, by categorizing the time parameter in terms of period of day, the day in the week, type of day (working day or sparing day), we can exploit the habit of the sequence of contexts. Group context parameters by restricting the boundary of contexts, is important to the specific environment. According to the analysis, the contexts are categorized into 3 groups: shared context, dynamic context and intermediate context which plays both shared context and dynamic context roles. The intermediate context can play both the dynamic part and a shared context part which links to another dynamic context or intermediate context. Furthermore, the context itself links directly to the action. The action in this case means the preferences of the devices and appliances responds to users' behaviors. In the latter part in the described scenario, the action "Take a rest" includes the state of the air condition, TV, and the light in the bedroom.

*B. Contextual Graph*

The basic idea of contextual graph relies on the fact that past contexts can be remembered and adapted to solve the current context. The context is managed to organize in the graph type. In the contextual graph, rather than creating a solution from scratch, the contexts similar to the current context are retrieved from memory. The best match is selected and adapted to fit the current context based on the differences and similarities between the two contexts.

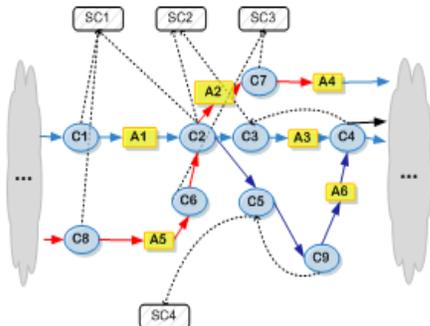

Fig. 1. An example of contextual graph

The contextual graph consists of nodes and paths connecting nodes. A node in the graph can be a context node or an action node. Fig. 1 shows an example about the proposed contextual graph. The opaque round shapes denote the *action node* whilst the rectangles stand for the *context nodes*. The *shared contexts* in the dashed rectangle connect with the action via a dotted line. It is easily realized that a context node has no or many arrows which link to its shared contexts. Every arrow has a direction from the start node to the end node. Furthermore, one path has a unique ID to distinguish from the others. Another issue often occurs in contextual graph is the context crossroad. The reason is that there are many *paths* going through a node and one specific context has many possible previous contexts. Therefore, the context crossroad occurs when the current context has many actions to be operated. The action node lies on the link connecting between the ex-context and the current context will be operated. The cloud shapes represent the rest of the context space. The graph can be also used as an input for user preferences machine learning.

IV. THE PROPOSED FRAMEWORK

This section describes the context-awareness framework for ubiquitous computing middleware by integrating the components for reasoning the contextual graph. The middleware framework with the whole system infrastructure does gather context information, process it and derive meaningful actions from it.

*A. Overall architecture*

The building blocks consist of the following 7 components: Stored Context Database, Graph Builder, Context Creator, Matching Module, Path Module, Rules Store and Context Reasoning Module. Fig. 2 shows the architecture overview of the framework described afterwards. The brief description on each of components is following.

The **context filtering module** selects essential contexts to be stored in the database. This module eliminated redundant or useless information received from the environment, people. Essential contexts are stored in the database for context learning. Before storing into the database, this module categorizes the contexts according to the relevance to users and context parameters. By saving the classified data through the filter module, the desired information can be obtained.

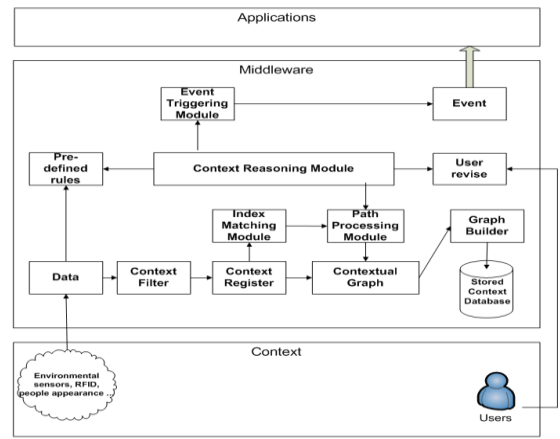

Fig. 2. Architecture overview

Whenever environments change, instances of context property also change. To recognize the change of environments, the application acquires events for property instances' change according to event lists. The **context creator** generates an application-specific context object and also registers this context with the context-aware system before logging into the database. The **rules store** saves all the rules activated when they match with the context object. The action will be retrieve at this step when the context satisfies one of the rules in the list. The rule existing in the list has the certainty such as "The light will be turned off if there is nobody in the house or in the office".

The ***stored context database*** stores all significantly changed context information in environments. The ***graph builder*** constructs contextual graphs based on the data collected from the context database. The ***index matching module*** is responsible for finding the matching context to the current context. If the new context instance already exists in the database, the system doesn't need to register or store this context in the database. The ***path processing module*** handles the operations related to path, such as adding or finding a path.

The ***user revise module*** collects the feedback from the users to adjust the appropriate action which suits the current context. In the context-aware system, the ***context reasoning module*** queries and reasons the suitable action for the current context based on the results getting from the path processing module, user revise and rules store. The result actions can be invoked automatically without user's intervention by the ***event triggering module***. This module fires the event to the applications with the decision inferred from the context reasoning module.

### B. Context reasoning process

The main process of a contextual graph based context-aware system involves the following steps. In order to support problem-solving, an initial memory of previous contextual information is required. Selected previous cases need to be indexed and collected in memory. Much of the knowledge needed for context-aware systems is in the set of contextual information. Collecting contextual information may not take as much. The storing memory must be organized into some manageable structure.

In the first step, an application initially describes its application context parameters and then registers the basic rules to the rules store in the system. The context creator next is responsible for creating the new context corresponding to the sensed information collected from various and diverse sensors. This context is applied with the rules store. After that, the matching module tries to find the best relevant context to the new context. The system recalls contextual cases that have relatively high similarity values, i.e., previous contextual cases with similar indexes are retrieved. Partial matching is conducted by employing evaluation functions to the match. The context retrieval techniques include fuzzy mathematical method, nearest neighbor search, and statistical weighting methods. In the case of not finding the matching context, the context and its solution should be added to stored context database for future reference and the graph is updated by the graph builder. On the contrary, if a similar context in the graph is found, the path processing module manages to find the path connecting to the corresponding action node.

Then, if there is an existing action connecting to the context, the retrieval action from graph is synthesized with the retrieval action from rules to conclude the final action by the contextual reasoning module. Otherwise, the new action, which is obtained by the user revise module or automatically updating the status of devices and appliances, is stored with the context if there is no action connecting to the context in the graph. Context-aware system then uses the gathered contextual knowledge to infer what the user expects and execute the expected services. The event triggering module fires the final event to the application meaning to provide the solution for the current context.

Contextual graph based context-aware systems allow the user to solve a problem by making use of *previous chain of experience*. Therefore, the context-aware system reasoning results can be the closest matches to actual human reasoning than rule-based knowledge or case-based knowledge representation. By applying this operation, pervasive application when receiving the new context asynchronously simply knows how to react based on the shared context and dynamic context in the contextual graph.

### V. APPLICATION USING THE PROPOSED FRAMEWORK

In this section, we show an application, uClassroom, in order to prove that our proposed contextual graph makes context modeling and reasoning easier and simpler.

### A. Scenarios

uClassroom is an application for classroom in a ubiquitous computing environment. Professor Lee has a Multimedia class in this uClassroom on Wednesday at 2 p.m. After Professor Lee checks in the classroom through RFID reader, the "prepare classroom" action has been called. The light is turned on and the projector starts showing the lecture file on the screen. Additionally, throughout the class, there are some exams, such as mid-term or final test. On the test day, the Teaching Assistant Park would enter the classroom first. After stabilizing the students, Professor Lee would enter the classroom to explain clearly all the questions related to the test. And then, he would leave the uClassroom. The uClassroom, instead of preparing the lecture as usual just as Prof. enters, changes to supervised mode by turning on the camera and start counting the testing time. When Prof. Lee gets out the room, the uClassroom doesn't turn into the finished mode. The finished mode "Close the classroom" is just only activated when the Teaching Assistant checks out the uClassroom instead.

### B. The operation of the proposed context-aware system

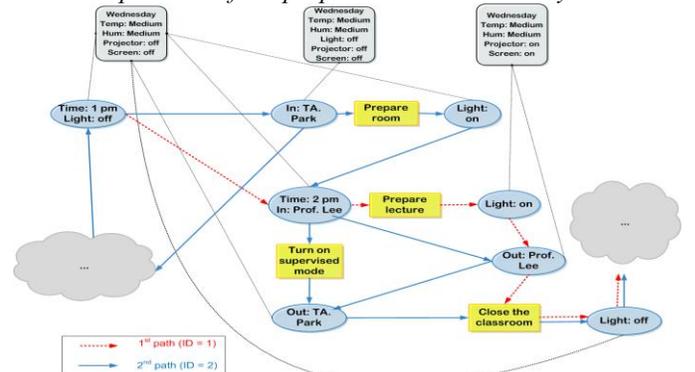

Fig. 3. Contextual Graph for uClassroom

As seen above, Fig. 3 illustrates the corresponding contextual graph for the scenario of uClassroom in terms of the steps of operation. Firstly, the sensed information is converted into context object by the context creator. On the test date, when TA Park enters the room, the path with ID = 2 is chosen and the corresponding action is "Prepare room" instead of "Prepare lecture". The new context of the room is "Light: on" with no change in the shared context, project and screen are off. The new context is created with 2 actions in the selection list when Professor Lee gets into the room. The context reasoner chooses the "Turn on supervised mode" because this action lies on the path has ID = 2. Similarly, another new context is created when Professor Lee gets out. However, the action "Close the classroom" is not called because this action isn't in the path with ID = 2. This action is just operated when TA Park checks out the room.

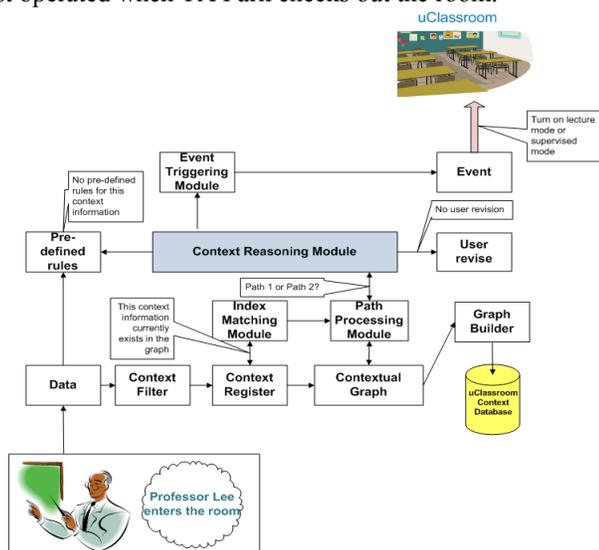

Fig. 4. The diagram of uClassroom system

As a part of our long term research plan, we are prototyping an intelligent room to demonstrate the feasibility of our approach. The context-awareness part of this uClassroom is implemented based on the proposed framework. We already controlled the appliances such as projector, the light mentioned in the scenario by using home appliances control system based on infrared ray and power line [14]. In the two-way communication, users can get the statuses of the appliances, and they can also set new statuses for them as well. In Fig. 4, the event is finally fired to the uClassroom system as the action "Turn on lecture mode" or "Turn on supervised mode". As a result, using contextual graph is a promising approach for the context-aware system.

VI. CONCLUSIONS AND FUTURE WORKS

This paper introduces a novel framework of context-awareness for ubiquitous computing based on Contextual Graph. The presented framework exploits some potentialities of contextual graphs. First, contextual graphs can use the previous knowledge to decide the action for the current context. It is clear that the more a system based on contextual graphs is used, the more it will preserve corporate memory. Second, contextual graphs deal with problems facing in other approaches like decision trees and case-based reasoning. The last but not the least is that contextual graphs construct the knowledge architecture for users' preferences learning. The proposed context-awareness framework and the context reasoning based on contextual graph are the main contributions of this paper. Graph-based systems present a powerful possibility to support knowledge exchange by visualizing shared context. Furthermore, knowledge transparency inside the organization is improved.

Our work shows that applying contextual graph is suitable for building a common knowledge representation for context-aware systems to share and reason with contextual knowledge. We expect that framework will help to overcome problems from context reasoning. Our future works focus on deploying more applications on the experiment room by using context-aware framework based on contextual graphs.

ACKNOWLEDGEMENT

This paper is supported by the Industry Promotion Project for Regional Innovation. The authors also would like to thank the BK21 Project of Korea and the anonymous reviewers for useful comments.